\documentclass[journal=jctc,manuscript=article]{achemso}

\usepackage[version=3]{mhchem} 
\usepackage{amssymb}
\usepackage{amsmath} 
\usepackage{url}

\author{Zijian Meng}
\email{contact@richardzjm.com; 17zjm1@queensu.ca}
\author{Karim Zongo}
\author{Matthew Thoms}
\affiliation{Department of Mechanical and Materials Engineering, Queen's University, Kingston, ON
Canada}
\author{Ryan Grant}
\affiliation{Department of Electrical and Computer Engineering, Queen's University, Kingston, ON
Canada}
\author{Laurent Karim Béland}
\affiliation{Department of Mechanical and Materials Engineering, Queen's University, Kingston, ON
Canada}

\title{Accelerating Moment Tensor Potentials through Post-Training Pruning}

\abbreviations{IR,NMR,UV}

\begin{document}







\begin{abstract}

Moment Tensor Potentials (MTPs) are machine-learning interatomic potentials whose basis functions are typically selected using a level-based scheme that is data-agnostic. We introduce a post-training, cost-aware pruning strategy that removes expensive basis functions with minimal loss of accuracy. Applied to nickel and silicon–oxygen systems, it yields models up to seven times faster than standard MTPs. The method requires no new data and remains fully compatible with current MTP implementations.

\end{abstract}


Machine-learning interatomic potentials (MLIPs) offer a tunable compromise between the accuracy of first-principles methods and the efficiency of empirical potentials. MLIP frameworks include neural network potentials, Bayesian approaches, and linear regression on systematically constructed basis functions. The latter category provides a practical balance between accuracy and efficiency, encompassing established formalisms such as the Spectral Neighbor Analysis Potential (SNAP)~\cite{thompson2015spectral}, the Moment Tensor Potential (MTP)~\cite{shapeev2016moment}, and the Atomic Cluster Expansion (ACE)~\cite{drautz2019atomic}.

Efforts to improve the efficiency of MLIPs have been made on several fronts. On the training side, one major direction has been to reduce the cost of generating high-quality data, which depends on expensive quantum-mechanical calculations. Active learning strategies~\cite{podryabinkin2017active,zhang2019active} and automated data-driven generation~\cite{poul2023systematic,karabin2020entropy} have lowered the cost of curating effective training datasets. Some of our recent work has shown that smaller training configurations can capture essential physics while substantially reducing data-generation costs~\cite{luo2023set,meziere2023accelerating,sun2024interatomic,meng2025small}.

As machine-learning potentials are increasingly applied in large-scale MD, the computational cost of evaluating atomic energies and forces can become a limiting factor. On the hardware side, several efforts have focused on high-performance implementations using GPU acceleration through Kokkos-based backends in the LAMMPS~\cite{thompson2022lammps} MD engine~\cite{nguyen2021billion,lysogorskiy2021performant}. We recently contributed an efficient GPU-enabled MTP implementation in LAMMPS~\cite{meng2025kokkos}.

A complementary direction for improving efficiency lies in the algorithmic design of the basis set itself. For example, Thompson \textit{et al.} used the DAKOTA optimization framework~\cite{adams2011dakota} to tune hyperparameters in SNAP, such as the bispectrum components and cutoff radius~\cite{thompson2015spectral}. In ACE, users specify hyperparameters such as the correlation order~($\nu$), the maximum radial index~($n_\text{max}$), and the maximum angular momentum~($l_\text{max}$), which together define the basis. The basis can then be systematically expanded using a hierarchical ladder fitting scheme \cite{lysogorskiy2021performant,bochkarev2022efficient}.

By contrast, the MTP basis is governed by a single level parameter that counts combinations of radial and angular indices with fixed weights. This design is straightforward and systematic, but it implicitly assumes that all tensor contractions contribute equally to the balance between accuracy and computational cost. In practice, the basis functions are evaluated through a recursive style compute tree that is highly cost-asymmetric: different contractions can differ in cost by more than an order of magnitude. Recently, Wang \textit{et al.} proposed alternative level definitions that increase the number of radial basis functions, leading to improved performance~\cite{wang2025efficient}.

Here, we introduce a post-training pruning strategy for the MTP basis that explicitly accounts for computational cost. Starting from a fully trained base potential, we apply a multiobjective evolutionary algorithm (EA) to identify subsets of basis functions that jointly optimize accuracy and cost. Accuracy is estimated by refitting the linear coefficients while keeping the radial components fixed, and cost is evaluated heuristically by traversing the compute tree. The procedure yields a family of potentials distributed along a cost–accuracy Pareto front. The resulting pruned MTPs can be used directly with existing implementations, including MLIP-2~\cite{novikov2020mlip}, MLIP-3~\cite{podryabinkin2023mlip}, and our LAMMPS implementation~\cite{meng2025kokkos}, without code modification. Although pruning was tuned for the CPU implementation, the resulting models remain compatible with GPU backends with potentially different scaling behavior. Additional details are provided in the Supporting Information.

The optimization workflow was implemented in Python using the \texttt{pymoo} framework~\cite{pymoo}, primarily employing the Non-dominated Sorting Genetic Algorithm II (NSGA-II)~\cite{deb2002fast}, with limited tests using the Multiobjective Evolutionary Algorithm based on Decomposition (MOEA/D)~\cite{zhang2007moea}. Numerical operations relied on \texttt{NumPy}~\cite{harris2020array}, and compute-tree traversal was accelerated with \texttt{Numba}~\cite{lam2015numba}. Parallel evaluations were performed using \texttt{mpi4py}~\cite{dalcin2005mpi,dalcin2021mpi4py}, replacing \texttt{pymoo}'s built-in parallel backend. While strong scaling remains bounded by serial components of the evolutionary algorithm, this approach enables efficient exploration of the Pareto front. The code is open-source and available as a pip-installable package from GitHub\cite{meng2025prune}.

\begin{figure}[htb]
    \centering
    \includegraphics[width=0.49\linewidth]{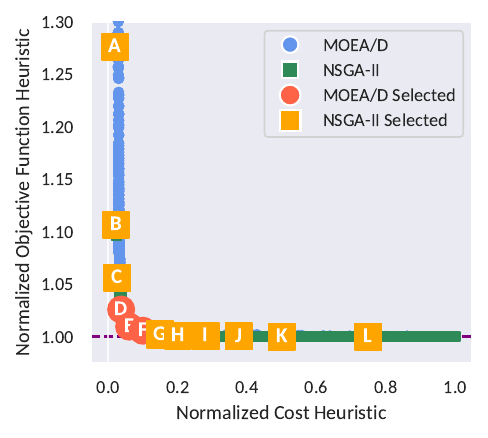}
    \hfill
    \includegraphics[width=0.49\linewidth]{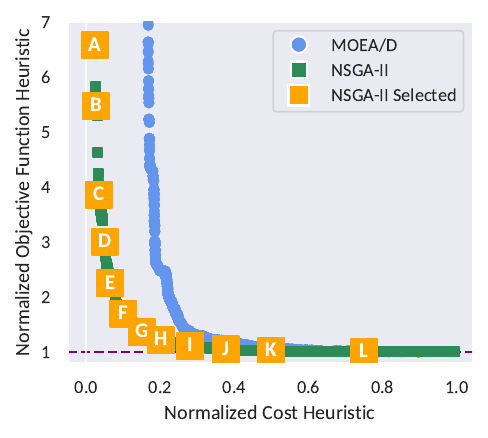}
\caption{\textbf{Left: Nickel. Right: Silicon–oxygen.} Cost–accuracy Pareto fronts obtained from the pruning process using NSGA-II and MOEA/D, each normalized to the base potential (MTP level~28). Twelve representative potentials, labeled~A–L, were selected to span the Pareto fronts and illustrate the range of achievable cost–accuracy tradeoffs. NSGA-II consistently identified models with lower computational cost and higher accuracy compared with MOEA/D.}

    \label{fig:heuristic}
\end{figure}

We demonstrate our pruning strategy on two representative systems: nickel, using the dataset of Andolina \textit{et al.}~\cite{andolina2023highly} from ColabFit~\cite{vita2023colabfit} as an example of a simple face-centered-cubic metal, and silicon–oxygen, using the 1159-structure dataset of Zongo \textit{et al.}~\cite{zongo2024unified} as a complex covalent–ionic system. For both materials, reference MTPs were trained using MLIP-3 across levels 6–28 following the standard level definitions of the MLIP code. Each level was fitted 32 times with randomized initial parameters, and the model with the lowest training loss was retained for further evaluation. The level-28 MTP served as the base potential for subsequent pruning. The resulting cost–accuracy Pareto fronts are shown in Fig.~\ref{fig:heuristic}, from which twelve representative potentials were selected, taking the better-performing results from either of the two evolutionary algorithms.

\begin{figure*}[htb]
    \centering
    \includegraphics[width=\linewidth]{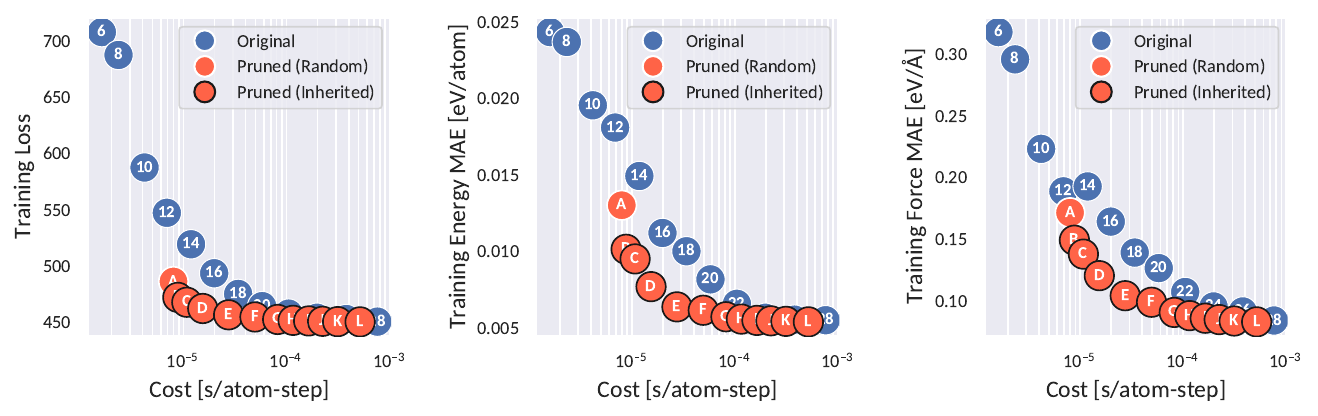}
    \centering
    \includegraphics[width=\linewidth]{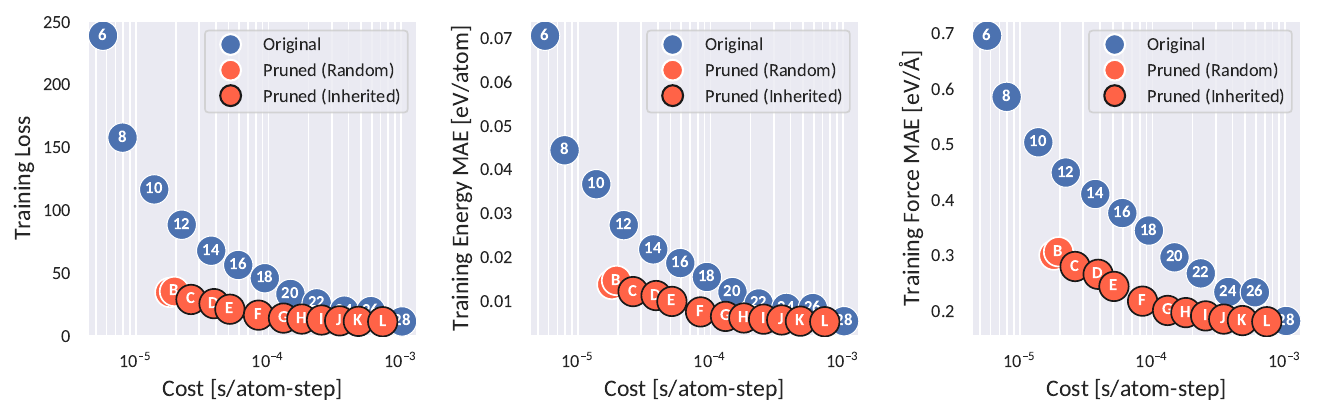}
\caption{\textbf{Top: Nickel. Bottom: Silicon–oxygen.} Cost–accuracy comparison between the original level-based MTPs and the pruned MTPs. Panels show (left) training loss, (middle) per-atom energy (MAE), and (right) force MAE. “Inherited” and “Random” denote the two initialization methods for fitting the pruned potentials. Computational costs were measured on a single AMD EPYC~9654 core using systems of 2048~FCC Ni atoms and 1944~$\alpha$-quartz atoms. Representative examples with comparable or improved accuracy show speedups of 3.8$\times$ (Ni, model~E) and 7.0$\times$ (Si–O, model~F) relative to level-based MTPs.}

    \label{fig:training_results}
\end{figure*}

To assess how initialization affects the fitting of the pruned potentials, we compared two strategies: random initialization and parameter inheritance from the base level-28 MTP. In the random initialization strategy, the twelve representative MTPs (A–L) were fitted from scratch using MLIP-3, following the same random seeding protocol as for the level-based potentials. The inherited initialization refitted MTPs (A-L) with MLIP-3 using the base potential's parameters as a starting point. Fig.~\ref{fig:training_results} compares the better-performing initialization for each pruned model against the original MTP levels in terms of training loss, energy per atom mean absolute error (MAE), and force MAE. In both material systems, inherited initialization generally yielded faster convergence and lower training error, except for the lowest-cost models, where both initialization schemes performed comparably.

In nickel, model~E achieves lower training loss and mean absolute errors for both energy and forces than the level-22 MTP, while running 3.8$\times$ faster. In silicon–oxygen, model~F attains better training accuracy than the level-26 MTP with a 7.0$\times$ reduction in cost, and model~A performs similarly to the level-20 MTP while being 8.1$\times$ faster. The speedups in Si-O are greater due to more opportunities for optimization. Nickel is a simpler system where we observed convergence in training loss around MTP level 22.


The structure of the pruned models clarifies the limitations of data-agnostic, level-based construction schemes. In Si–O, for instance, the level-12 MTP contains 127 learnable parameters, whereas potential~A is 22.9\% faster despite using 386 parameters. The discrepancy arises from the allocation of learnable parameters rather than their count. The MTP includes per-neighbor and per-neighborhood costs, and Si–O has many neighbors on average. Thus, the pruning retains basis functions that reduce per-neighbor costs, whereas Ni, with only 44\% of Si–O's average neighbors, benefits from retaining them. For instance, Potential L in both case studies achieves a $\approx$40\% speedup relative to their base potentials without compromising fit quality. However, in Ni, all per-neighbor costs are retained, while in Si-O, 17\% are pruned. Additionally, all pruned potentials retained all the base potential's radial basis sets, corroborating Wang \textit{et al.}'s alternative level definition, which promotes more radial basis sets.

\begin{figure*}
    \centering
    \includegraphics[width=0.95\linewidth]{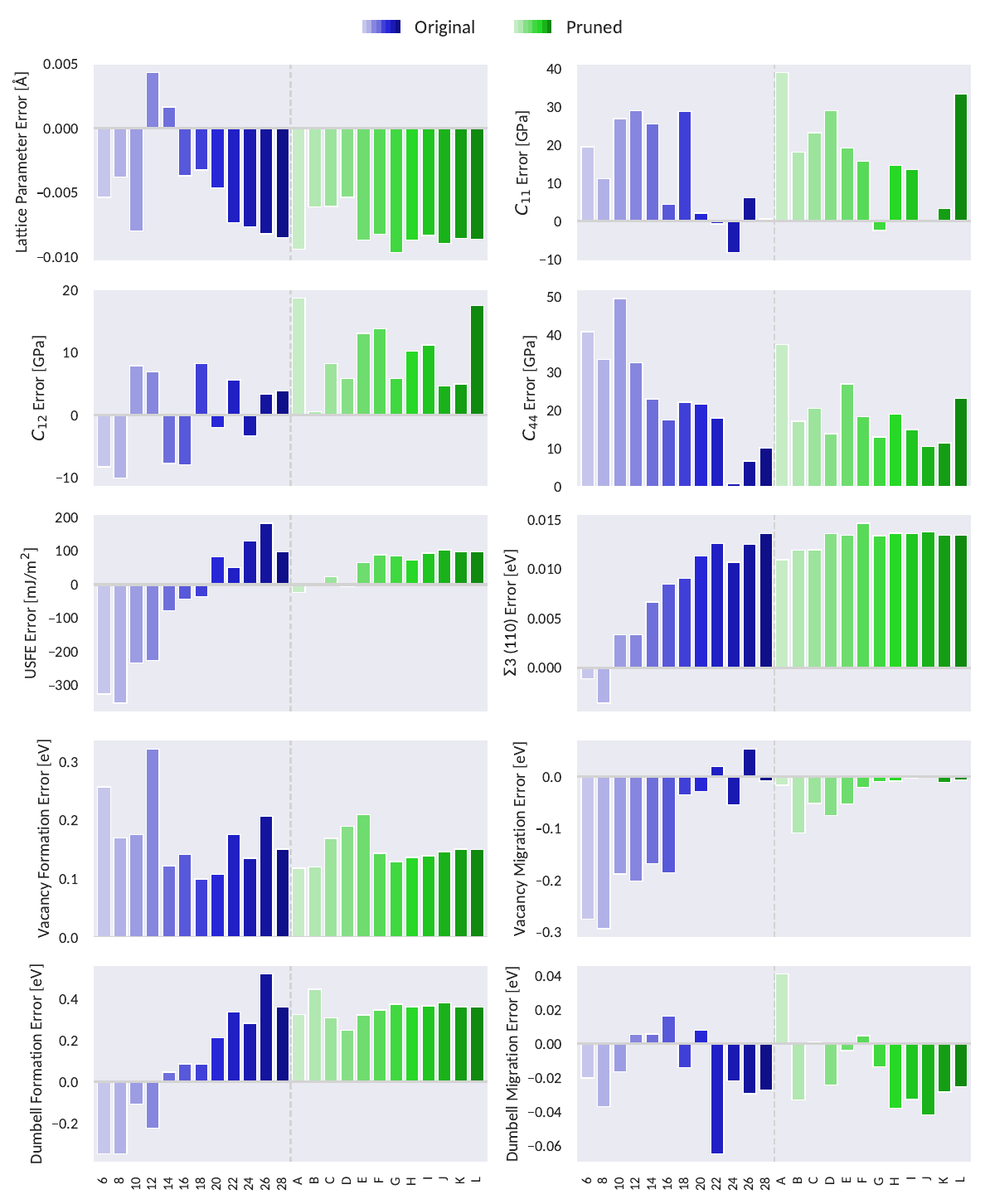}
\caption{\textbf{Nickel MTP errors.} Panels show, from top to bottom: lattice parameter; elastic constants $C_{11}$, $C_{12}$, and $C_{44}$ at 300~K; unstable stacking fault energy along $\langle 112 \rangle$ on the $\{111\}$ plane; $\Sigma3(110)$ tilt grain boundary formation energy; and formation and migration energies for monovacancies and $\langle 110 \rangle$ dumbbell self-interstitials. Pruned potentials exhibit similar error to the level-28 MTP, indicating that pruning primarily removes inefficient basis functions without greatly altering the fitted behavior. Color luminance indicates models within each group ordered by increasing cost.}

    \label{fig:ni_props}
\end{figure*}
\begin{figure*}
    \centering
    \includegraphics[width=0.8\linewidth]{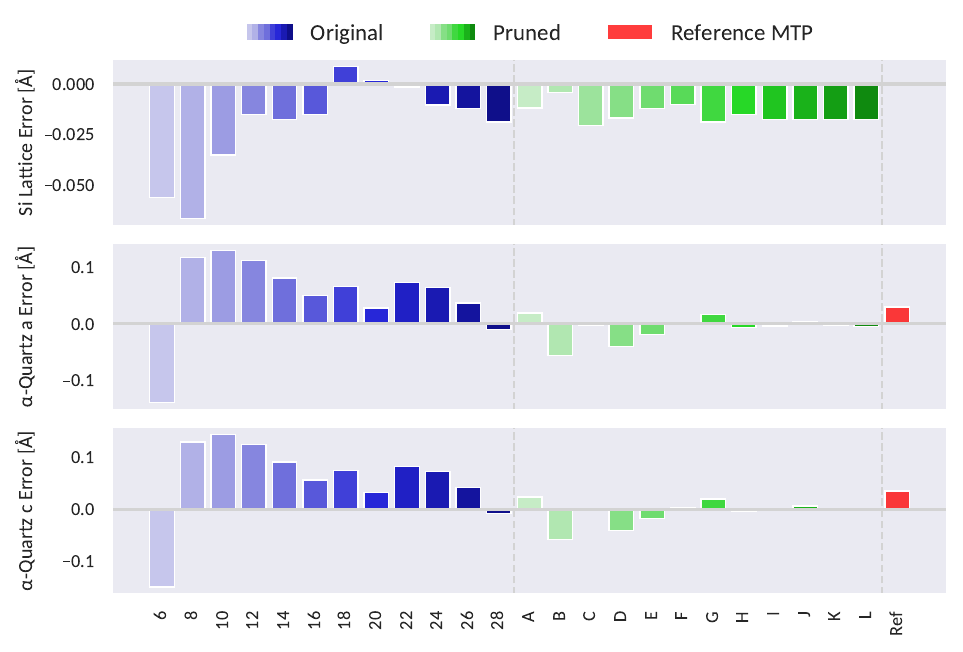}
\caption{\textbf{Top to bottom:} silicon lattice parameter, $\alpha$-quartz $a$, and $\alpha$-quartz $c$. Errors relative to DFT at 0~K are shown for the original level-based MTPs, the pruned models, and the reference level-26 potential from Zongo~\textit{et~al.}~\cite{zongo2024unified}. The reference potential was trained with different hyperparameters and initialization settings from those used here. Pruned models reproduce these equilibrium properties with comparable accuracy while achieving substantial reductions in computational cost. Color luminance indicates models within each group ordered by increasing cost.}

    \label{fig:si_lattice}
\end{figure*}
\begin{figure*}[h!]
\centering
{\setlength{\abovecaptionskip}{3pt}%
 \setlength{\belowcaptionskip}{0pt}%
 \setlength{\textfloatsep}{5pt}%

\hspace{\fill}
\includegraphics[width=0.38\linewidth]{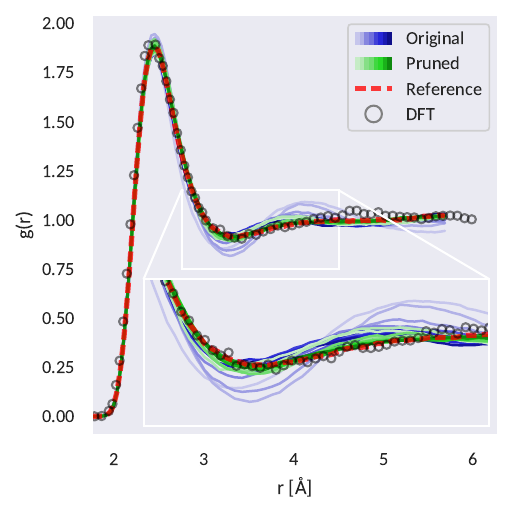}%
\hspace{0.05\linewidth}%
\includegraphics[width=0.38\linewidth]{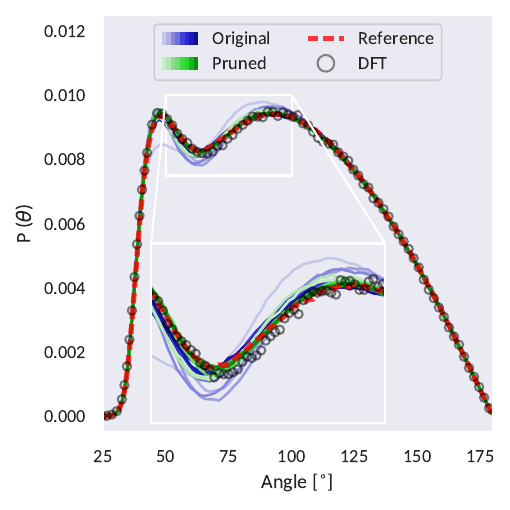}%
\hspace{\fill}

\vspace{-7pt}

\hspace{\fill}
\includegraphics[width=0.38\linewidth]{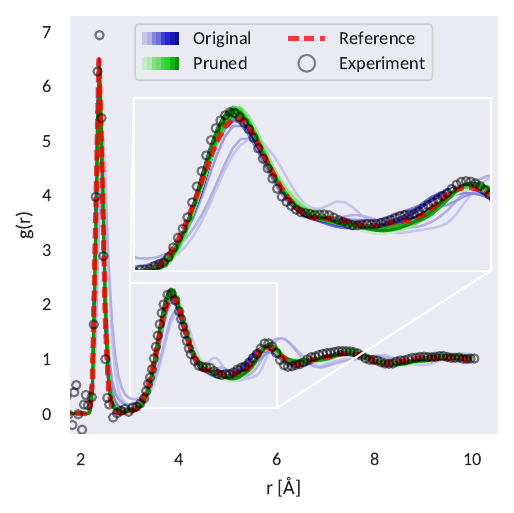}%
\hspace{0.05\linewidth}%
\includegraphics[width=0.38\linewidth]{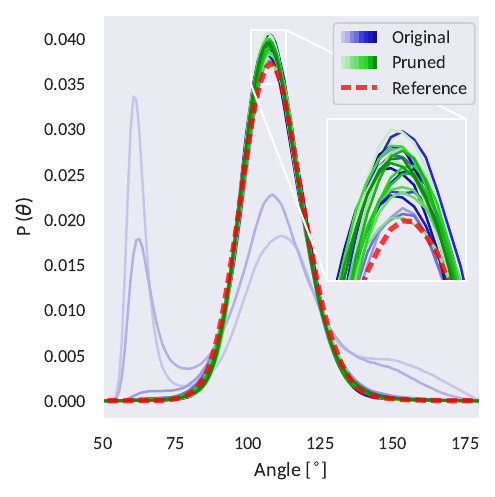}%
\hspace{\fill}

\vspace{-7pt}

\hspace{\fill}
\includegraphics[width=0.38\linewidth]{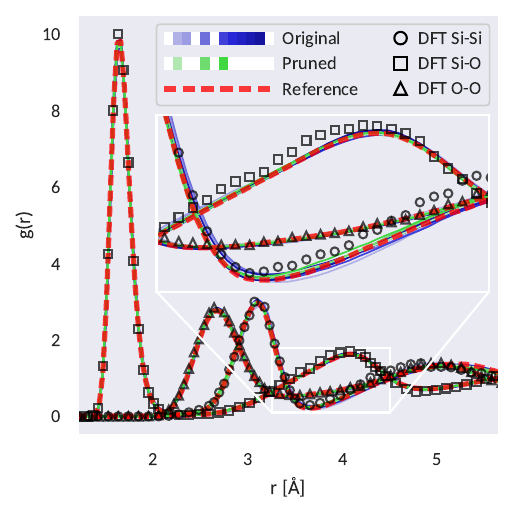}%
\hspace{0.05\linewidth}%
\includegraphics[width=0.38\linewidth]{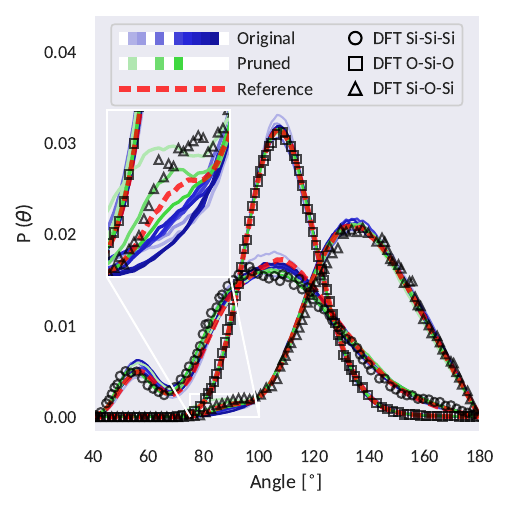}%
\hspace{\fill}

\caption{\textbf{Top to bottom:} liquid Si, amorphous Si, and liquid SiO$_2$. RDF (left) and ADF (right) are shown for the original level-based MTPs and the pruned models. DFT results and the reference level-26 MTP from Zongo~\textit{et~al.}~\cite{zongo2024unified} are included for comparison. Experimental data is from Laaziri~\textit{et~al.}~\cite{laaziri1999high}. Insets highlight regions of higher variability. Pruned models reproduce the main DFT and experimental features across all systems while substantially lowering computational cost. Color luminance indicates models within each group ordered by increasing cost; in silica, white denotes unstable potentials.}

\label{fig:dfs}}%
\end{figure*}

To assess whether the accuracy gains observed during training are reflected in physically meaningful quantities, we compared the pruned and level-based MTPs across a range of physical properties. The complete benchmark data and trained potentials are available from Zenodo~\cite{meng_2025_accelerating}. In nickel, the benchmarking suite of Thoms~\textit{et~al.}~\cite{thoms2025benchmarking34openkimnickel} was used for Fig.~\ref{fig:ni_props}, including the equilibrium lattice parameter, 300~K elastic constants, the unstable stacking fault energy (USFE) along $\langle 112 \rangle$ on $\{111\}$, the $\Sigma3(110)$ tilt grain boundary formation energy, and the formation and migration energies of monovacancies and $\langle 110 \rangle$ dumbbell self-interstitials. In silicon–oxygen, results were compared against DFT and the level-26 MTP distributed alongside the Zongo~\textit{et~al.} dataset~\cite{zongo2024unified}. That reference potential employed different fitting hyperparameters targeting the accuracy of structures and was initialized randomly, with the best seed selected based on physical-property performance rather than training loss. Fig.~\ref{fig:si_lattice} presents the Si and $\alpha$-quartz lattice parameters, and Fig.~\ref{fig:dfs} shows the radial (RDF) and angular (ADF) distribution functions for liquid silicon, amorphous silicon, and liquid silica.

Overall, the pruned models reproduce the behavior of the full level-28 potential remarkably well. This is especially true for models retrained with inherited initialization. In other words, pruning alters the computational workload far more than it alters the underlying physics captured by the fit. The main exception is the nickel elastic constants, which show slightly larger variations—a reminder that elastic responses are often among the most sensitive quantities in any interatomic potential.

In liquid silica at 3000~K, several potentials became unstable during molecular dynamics, an outcome that depended strongly on the initialization. By using different random seeds, both the level-based and pruned potentials could be stabilized, suggesting the instability of the pruned potentials may have been inherited from the unstable base potential. The instabilities likely arise from nonphysical behavior near the MTP's minimum distance. An additional short-range repulsive model, such as a Ziegler-Biersack-Littmark (ZBL) \cite{ziegler1985stopping} or Nordlund-Lehtola-Hobler (NLH) \cite{nordlund2025repulsive} potential—together with MTP smoothing at the transition—would likely eliminate these issues. A more subtle example of inherited behavior appears in the Si–O–Si angular distribution of liquid silica (Fig.~\ref{fig:dfs}, inset, bottom right). Between 80° and 100°, DFT predicts a secondary peak absent in the original level-based potentials. Interestingly, every stable pruned potential reproduces this feature to some extent, implying that pruning preserved some information encoded in the more complex base potential.

Overall, our approach shows that substantial performance gains can be achieved in MLIPs by optimizing their structure. Even with our simple heuristics for estimating accuracy and cost, the resulting potentials offer strong speedups. In practice, pruning can be tuned to match specific simulation goals. For instance, the neighbor count used in our pruning was the average of the training set, but could easily be adapted to high-temperature or high-density conditions. Furthermore, pruning begins from a rich base potential; it compresses what exists rather than adding new terms. Starting from larger bases, such as MTPs with more radial basis sets, would open even broader optimization spaces, although higher pruning costs would be incurred.

While we have focused on the MTP, a similar approach could be considered in other formalisms. ACE may be a good candidate for cost-aware optimization since it uses a similar recursive style evaluation of its basis functions.

From a practical standpoint, pruned MTPs can be used immediately in MLIP-3 and our own LAMMPS MTP module, requiring no code changes and yielding speedups of up to sevenfold. Faster potentials enable longer simulations, larger systems, and broader statistical sampling on the same hardware. Because our post-training method requires no new training data and is released as an open-source package with examples, it can be easily adopted by the community.

\begin{acknowledgement}

This work was funded by the University Network for Excellence in Nuclear Engineering (UNENE), the Natural Sciences and Engineering Research Council of Canada (NSERC), and Mitacs. We thank the Digital Research Alliance of Canada (DRAC) for the generous allocation of computer resources. 


\end{acknowledgement}

\begin{suppinfo}
Calculation methodology of heuristics, pruning evolutionary algorithm details, and silicon-oxygen physical property calculation details (PDF). 

\end{suppinfo}

\section*{Declaration of Generative AI Use}

Generative AI tools were used only for grammar, wording, and spell checking; accelerating code prototyping, commenting, and refactoring (all ideas and algorithms are the authors’ own); writing Python scripts to generate plots from our own data; and correcting \LaTeX{} formatting and equation issues. All intellectual and analytical contributions are solely those of the authors.

\bibliography{main}

\end{document}